\documentclass[aps,preprint,preprintnumbers,nofootinbib,showpacs]{revtex4}

\def\beq{\begin{equation}}
\def\eeq{\end{equation}}
\def\beqa{\begin{eqnarray}}
\def\eeqa{\end{eqnarray}}

\def\lsim{\mathrel{\raise.3ex\hbox{$<$\kern-.75em\lower1ex\hbox{$\sim$}}} }
\def\gsim{\mathrel{\raise.3ex\hbox{$>$\kern-.75em\lower1ex\hbox{$\sim$}}} }

\begin{document}
\draft
\preprint{\tighten{\vbox{\hbox{NCU-HEP-k010}\hbox{Aug 2003}
}}}
\vspace*{1in}

\title{\bf Anomaly Free Gauged $SU(4)_L\times U(1)_X$ Models with Little Higgs
\vspace*{.3in} }
\author{\bf Otto C. W. Kong \vspace*{.2in}}
\email{otto@phy.ncu.edu.tw}
\affiliation{
Department of Physics, National Central University, Chung-li,
TAIWAN 32054  \\
\& \ \ \ Institute of Physics, Academia Sinica, Nankang, Taipei, Taiwan 11529
\vspace*{.5in}    	}

\begin{abstract}
We present an analysis showing how anomaly free fermionic spectra with  consistent 
embeddings of the Standard Model spectrum under $SU(3)_C\times SU(2)_L\times U(1)_Y \; 
\subset \; SU(3)_C\times SU(N)_L\times U(1)_X$ for any $N>2$ can be obtained,
with special focus on the $N=3\,\mbox{and}\,4$ cases. The construction is motivated
by the little Higgs mechanism. We discuss the relevancy of the fermionic spectra to the latter,
concentrating on two $N=4$ models, without fermions of exotic charges. Such models hold 
the promise to address and solve all the major theoretical as well as phenomenological 
problems of the  Standard Model at the TeV scale.
\end{abstract}
\maketitle

The phenomenologically overwhelmingly successful Standard Model (SM) has a few
theoretically shortcomings. The latter motivates many high energy theorists looking
for beyond SM structures that could help us understand basically why some of the 
parameters in the SM have the values they have. The Higgs sector parameters, in particular,
are haunted by the hierarchy problem, or the very un-natural fine-tuning required due to 
the quadratically divergent quantum corrections to the Higgs boson mass. The fine-tuning
problem can be alleviated, only if there is new physics at the TeV scale that guarantee
the cancellation of the quadratic divergence to an acceptable level, or totally change our
picture of SM physics. A guaranteed cancellation has to come from some mechanism 
protected by a symmetry. Candidates of the kind include supersymmetry, and the recently
proposed little Higgs mechanism\cite{dd,S}. The fermion sector also has a very puzzling
flavor problem. Finally one may worry about the strong CP problem\cite{scp}. 

To appreciate the flavor problem, we want to emphasize first that the SM has fermions
of three families each has the same set of quantum numbers intricately connected among 
the family by chiral gauge anomaly cancellation conditions. This is well illustrated by the
following argument. Assuming the existence of a multiplet having nontrivial charges under 
each part of the $SU(3)_C\times SU(2)_L\times U(1)_Y$ SM gauge group and asking 
simply for the minimal chiral spectrum with all the gauge anomaly canceled, one would 
arrive at the spectrum of a SM family as the only solution\cite{unc67}. So, the first 
fundamental question of the flavor problem is why there are three families, instead of one. 
The next step in the direction would be to understand the values of the set of flavor 
parameters and the strong hierarchy among them. In this aspect, there is a notion related 
to the hierarchy problem (of the Higgs sector parameters). Namely, only the third family 
fermions may have a significant role to play in the latter problem. The first two families 
are simply coupled to the Higgs boson too weakly. This is suggestive that the third family 
might be different from the first two.

The idea of connecting the three families through gauge anomaly cancellations of 
an extended gauge group is related to many interesting 
models\cite{unc67,SVS,331,FLT,unifi,unc}. For the benefit of the present perspective, 
we note the following two approaches. Firstly, Ref.\cite{unc67} treats the three 
families on the same footing while trying to duplicate the structure of the one-family 
SM spectrum (as presented above) for a three-family embedding generic 
$SU(N)\times SU(3)\times SU(2)\times U(1)$ gauge symmetry ($N=4$ or otherwise). 
The existence of family-changing gauge bosons says that such models could not 
be relevant till a scale of about $200\,$TeV. Treating the third family (quarks) 
different, Refs.\cite{SVS,331,FLT} have $SU(3)_C\times SU(3)_L\times U(1)_X$
models that leave $SU(2)_L$ singlets mostly as $SU(3)_L$ singlets. Arguably, 
such model spectra are not particularly appealing, though they do tie the three
families together. In fact, as we will show below, the two models are only examples 
of a one-parameter class of infinite number of models. However, such a model has a 
scale of relevancy at about a TeV. None of these extended fermion models 
addresses the hierarchy problem, though one could always incorporate 
supersymmetry  other wise.

On the other hand, the recent little Higgs model-building game has not been paying 
quite enough attention to the fermion sector and the all important issue of gauge 
anomaly cancellations. Looking at it as a effective field theory at the TeV scale without 
bothering  about the strong dynamics behind it, a little Higgs model needs at least an 
extra top-like quark with some appropriate (global) symmetry to cancel the quadratic 
divergence from the top itself. Extra gauge bosons may also be needed to cancel the 
quadratic divergence from the electroweak gauge bosons. So, a little Higgs model 
does have an extended gauge symmetry with extra fermions. To figure out the flavor 
structure of such a model, one does need to know the full quantum numbers of the 
SM fields. As pointed out in our earlier paper\cite{009}, the construction of the full 
fermion spectrum is nontrivial and unavoidable. A simple family universal 
embedding typically leaves nonvanishing gauge anomalies. Cancelling the latter
by adding extra fermionic multiplets does not guarantee that no phenomenologically
unacceptable extra SM chiral fermion would be introduced. To have a complete
and consistent model, we are forced to address the flavor question at the same time.
Consistent model solutions then hold promise to be models 
that address all the major theoretical, as well as phenomenological, problems
of the SM at an very accessible energy scale. Such a model spectrum typically also 
contains extra $SU(2)_L$ singlet neutrino states, and new interactions involving the 
SM neutrinos. Hence, it also provides a bonus for understanding neutrino physics 
at the same low energy scale. It is our focus here to look into relevant fermionic
spectra of the type that do have the consistent gauge anomaly cancellations.

Following Ref.\cite{009}, we take the group theoretically simple little Higgs model(s)
introduced in Ref.\cite{KS} and try to constructed compatible anomaly free fermionic 
spectra to complete the consistent TeV scale models, without bothering about the
strong dynamics suggested to be behind the picture at the scale of tens of TeV. 
Ref.\cite{KS}  starts with a model with as $SU(3)_C\times SU(3)_L\times U(1)_X$
gauge symmetry, the consistent fermionic spectrum of which we discussed in 
Ref.\cite{009}. The  $SU(3)_L$ model has a difficulty on the Higgs quartic coupling.
The latter motivated the authors to extend the little Higgs construction to a model with an
$SU(3)_C\times SU(4)_L\times U(1)_X$ gauge symmetry\cite{KS}. Here, we present
an analysis showing how anomaly free fermionic spectra with  consistent 
embeddings of the SM spectrum under $SU(3)_C\times SU(2)_L\times U(1)_Y \; 
\subset \; SU(3)_C\times SU(N)_L\times U(1)_X$ for any $N>2$ can be obtained,
with special focus on the $N=3\,\mbox{and}\,4$ cases. By a consistent embedding,
we mean that the full fermion spectrum does yield the three families SM fermions
together with extra states that are vectorlike after the gauge symmetry is broken
to that of the SM.  While such models can be considered to be of interest in their
own right, those among them that could serve as little Higgs models are considered
to be more interesting. After presenting the fermionic spectra construction, we will
take a look at such little Higgs models.

As said above, a little Higgs model typically has an extra top-like quark $T$ to cancel
the quadratic divergence from the top itself. The way to do it as presented in 
Ref.\cite{KS} is to extend the $t$-$b$ quark doublet to a fundamental representation
of an $SU(N)_L$ multiplet with $T$ included. Take the $N=4$ example.
The $t$-$b$ quark doublet of $SU(2)_L\times U(1)_Y$ is to be embedded into a 
$SU(4)_L\times U(1)_X$ quadruplet $Q^a$ as follows
\beq
4_{\scriptscriptstyle L} = \left(
t^a \quad b^a
 \quad T^a  \quad T'^a
\right)^T
\eeq
with an appropriate ${X}$-charge denoted by $X_{\scriptscriptstyle Q}$. 
The third state is the top-light quark $T$, with the usual electric 
charge of ${\mathcal Q}=\frac{2}{3}$; and $a$ represent the $SU(3)_C$ index.  Here, 
we keep the identity of the $T'$, {\it i.e.} its electric charge, unspecified for the moment. 
The vectorlike QCD spectrum is to be recovered by introducing the Dirac partners in 
$SU(4)_L\times U(1)_X$ singlets as
\beq \label{Qa}
1_{\scriptscriptstyle L} =  \bar{b}_a\;,
\quad  \bar{t}_a\;,
\quad  \bar{T}_a\;,
\quad  \bar{T'}_{\!\!a}\;.
\eeq
Their electric charge is to be given by their $X$-charge directly. The requirement amount to
nothing other than a specific choice of $X$-charge normalization. 

We are about to start on the discussion on constructing the anomaly free fermionic spectra.
We want to emphasize here the construction we will discuss is generic. To be explicit, we will
first relax the little Higgs mechanism requirement, {\it i.e.} we do not require the $Q^a$
quadruplet to contain the $T$ quark with electric charge $\frac{2}{3}$. We will return to
models of interest in view of the little Higgs mechanism discussed in Ref.\cite{KS} afterwards.

The basic approach in the kind of model-building exploits the fact that
\[
N_c = N_f \;;
\]
namely, the number of SM families (of fermions) $N_f$ happens to coincide with the number
of colors. While the extension of $SU(2)_L$ doublets into complex (anti-) fundamental 
representations $N_{\scriptscriptstyle L}$ or  $\bar{N}_{\scriptscriptstyle L}$ of  $SU(N)_L$ 
introduces nontrivial  $[SU(N)_L]^3$ gauge anomaly, if the three families of quark doublets 
are embedded into one $N_{\scriptscriptstyle L}$  and two $\bar{N}_{\scriptscriptstyle L}$ 
representations, only one net colored-$\bar{N}_{\scriptscriptstyle L}$ multiplet is left to 
contribute to the anomaly which may then be canceled by putting the three families of leptonic 
doublets into $N_{\scriptscriptstyle L}$'s. We take $N=4$ here for an explicit illustration. 
Denote the nontrivial $SU(4)_L$ multiplets, apart from 
$Q^a$, by $Q'^a_k=\bar{4}_{\scriptscriptstyle L}$
 ($k=1\;\mbox{or}\;2$) and $L_i={4}_{\scriptscriptstyle L}$ ($k=1\;\mbox{to}\;3$), and 
their $X$-charges by $X_{\scriptscriptstyle Q'}$ and $X_{\scriptscriptstyle L}$ respectively. 
We have then 
\beq
X_{\scriptscriptstyle Q'} = \frac{1}{3} - X_{\scriptscriptstyle Q}
\eeq
and
\beq
X_{\scriptscriptstyle L} = X_{\scriptscriptstyle Q} -\frac{2}{3} 
\eeq
from the requirement for correct embedding of the $SU(2)_L$ doublets, or getting the right
electric charge or hypercharge differences. In fact, we can take the electric charge 
embedding as given by 
\beq \label{Q}
{\mathcal Q} = \frac{1}{2} \, \lambda^3 
 + \frac{A}{3} \, \lambda^8 	
+\frac{B}{6} \, \lambda^{15} + X \;,
\eeq
with the normalization $\rm{Tr}\{\lambda^a\lambda^b\}=2\delta^{ab}$.
The correct doublet embeddings require 
\beq \label{abx}
A+B+X_{\scriptscriptstyle Q}=\frac{1}{6}
\eeq
besides the given relationship among the $X$-charges. The latter gives
$X_{\scriptscriptstyle Q}+2\,X_{\scriptscriptstyle Q'}+X_{\scriptscriptstyle L} =0$, 
an equation that also guarantees the cancellation of $[SU(4)_L]^2 U(1)_X$ gauge 
anomaly (provided that $N_c = N_f$). 

So far in our construction of the fermion spectrum, we are left with the freedom to specify
$X_{\scriptscriptstyle Q}$ and $A$ (or equivalently $B$). Interestingly, we do not need to
specify their values in order to obtain the full anomaly free spectrum if all one cares is 
to embed the full SM spectrum while allowing extra vectorlike pairs of $SU(2)_L$ singlet 
quarks and leptons of arbitrary electric charges. All we have to do is to finish the spectrum 
discussed so far with $SU(4)_L$ singlets required for the vectorlike pairings of all the 
fermions at the QCD and QED level. And this works essentially in the same way for the 
other $N$ values. There could be some redundancy in the full spectrum so obtained. Say, 
one does not need to put in Dirac partners for the SM neutrinos; singlets with no 
$X$-charges are as good as not being there; and vectorlike $SU(4)_L$ singlet lepton 
pairs that come up may be removed. To convince
the readers that the anomaly cancellations do work, let us outline the mathematics. The
$[SU(3)_C]^3$ anomaly vanishes as QCD is kept vectorlike. The kind of embedding by
putting all Dirac partners of states in nontrivial $SU(4)_L$ representations as singlets,
hence with ${\mathcal Q}=X$, also guarantee cancellation of the $[SU(3)_C]^2 U(1)_X$ 
gauge anomaly family by family. For the  $[grav]^2 U(1)_X$ anomaly, the trace of
$U(1)_X$ charges for the quarks within each family is obviously just a scale factor
different from that of $[SU(3)_C]^2 U(1)_X$. We are hence left only with the leptonic
contributions, which again cancel for each family. It is then obviously that the 
$[U(1)_X]^3$ anomaly has to vanish too --- an explicit checking of the algebra could 
also be done.

We have illustrated above essentially the existence of infinite number of SM embeddings 
into $SU(3)_C\times SU(N)_L\times U(1)_X$ ($N=4$ or otherwise) that could be 
phenomenologically viable.  For the $N=4$ case in particular, we still have the freedom
to choose $X_{\scriptscriptstyle Q}$ and $A$ as we like.
The discussion is presented in such a way that it is easy to see how similar constructions
would work for any other $N$. For example, we can take $N=3$ and $B=0$, to remove the
inadmissible $\lambda_{\scriptscriptstyle L}^{15}$ in Eq.(\ref{Q}). Choosing 
$X_{\scriptscriptstyle Q}=\frac{1}{3}$ then has $A=\frac{-1}{6}$ as fixed by Eq.(\ref{abx})
giving the only extra quark in $Q^a$ as $T$ (electric charge $\frac{2}{3}$) and essentially
the {331} little Higgs model as constructed in Ref.\cite{009} (see also Refs.\cite{SVS,FLT}). 
We illustrate here again the full fermion spectrum in Table I. 
In fact, the (first) extra quark state in $Q^a$ always has 
${\mathcal Q}= X_{\scriptscriptstyle Q}-2A+(B)=\frac{1}{6}-3A$ [{\it cf.} Eq.(\ref{abx})].
One can freely choose the value for this electric charge. The choice here fixes the value of
$A$, which in turn fixes that of $X_{\scriptscriptstyle Q}$ through Eq.(\ref{abx}). Say, 
picking ${\mathcal Q}=\frac{5}{3}$ gives $A=\frac{-1}{2}$ and
$X_{\scriptscriptstyle Q}=\frac{2}{3}$  yielding the fermion spectrum of Ref.\cite{331}. 
Hence, there is a one parameter class of such {331} models each containing a different 
extra (exotic) singlet quark. Only the one with exactly the $T$ quark, as discussed in 
Ref.\cite{009}, is relevant as a little Higgs model.  For $N>4$, there have to be 
extra parameters similar to $A$ and $B$ in an extended Eq.(\ref{Q}). The structure
of such anomaly free models are otherwise easy to appreciate.

Now, we focus on $SU(4)_L\times U(1)_X$ models with an interest in their compatibility
with the little Higgs idea. We first note that the two extra quarks in $Q^a$ have 
electric charges $X_{\scriptscriptstyle Q}-2A+B (=\frac{1}{6}-3A)$ and 
$X_{\scriptscriptstyle Q}-3B$,
according to Eq.(\ref{Q}). We are free to choose their values, which in turn fixed the
full anomaly free spectrum of a model. The little Higgs mechanism, as discussed in
Ref.\cite{KS}, requires a top-like $T$ quark with ${\mathcal Q}=\frac{2}{3}$. Doing this
fixes $A=\frac{-1}{6}$. Ref.\cite{KS} suggests taking the fourth quark in $Q^a$ as having 
the same electric charge. Further imposing that is equivalent to taking  
$X_{\scriptscriptstyle Q}=\frac{5}{12}$ and $B=\frac{-1}{12}$. Following our discussion 
above, one can easily obtain the full fermion spectrum (see Table II for an illustration).
However, as the last quark of $Q^a$ suggested in Ref.\cite{KS} does not play a role in the 
quadratic divergence cancellation, the choice seems to be arbitrary. Giving the wide range 
of possible alternatives consistent spectra
\footnote{It may be of interest to note that one such $SU(4)_L\times U(1)_X$ model has 
actually been available in the literature\cite{FLT}. The model has the two extra quarks
having electric charges $\frac{2}{3}$ and $\frac{5}{3}$, hence is like a combination of
the two available $SU(3)_L\times U(1)_X$ models\cite{SVS,331,FLT}. 
},
we have to consider more seriously about the question of picking the one that is most
desirable from the theoretical point of view. Moreover, apart from the choice of Ref.\cite{KS},
there is another choice which looks a bit special---
$X_{\scriptscriptstyle Q}=\frac{1}{6}$,  $A=\frac{-1}{6}$, and $B=\frac{1}{6}$, giving an
extra bottom-like $B$ quark together with the $T$ quark. This is the second model
spectrum we think may also be of special interest from the little Higgs perspective. 
The full fermion spectrum of the model is given in Table III.

For the little Higgs part of such models, let us see what we have from Ref.\cite{KS},
which is fully compatible with the fermion spectrum of Table II.  Four scalar quadruplets 
of the same quantum numbers are considered. The quantum numbers are such 
the quadruplets (or rather their conjugates) couple to the $Q^a$ quadruplet and the
$\bar{t}_a$ or $\bar{T}_a$ singlet. Under our notation here, the 
$4_{\scriptscriptstyle L}$ scalars have $X$-charges given by 
$X_{\scriptscriptstyle Q}-\frac{2}{3} (=\frac{-1}{4})$. These are nonlinear  sigma 
model fields that may be parametrized as [{\it cf.} Eqs.(50,51) of Ref.\cite{KS}]
\beqa
&& \Phi_1= e^{\scriptscriptstyle +i {\mathcal H}_{\! u} \frac{f_2}{f_1}}
\left( \begin{array}{c}
0 \\ 0 \\ f_{\!\scriptscriptstyle 1} \\ 0
\end{array} \right)
\qquad
\Phi_2= e^{\scriptscriptstyle -i {\mathcal H}_{\! u} \frac{f_1}{f_2}}
\left( \begin{array}{c}
0 \\ 0 \\ f_{\!\scriptscriptstyle 2} \\ 0
\end{array} \right)
\nonumber \\ \label{phi}
&& \Psi_1= e^{\scriptscriptstyle +i {\mathcal H}_{\! d} \frac{f_4}{f_3}}
\left( \begin{array}{c}
0 \\ 0 \\ 0 \\ f_{\!\scriptscriptstyle 3} 
\end{array} \right)
\qquad
\Psi_2= e^{\scriptscriptstyle -i {\mathcal H}_{\! d} \frac{f_4}{f_3}}
\left( \begin{array}{c}
0 \\ 0 \\ 0 \\ f_{\!\scriptscriptstyle 4} 
\end{array} \right) \;,
\eeqa
where ${\mathcal H}_{\! u}$ and  ${\mathcal H}_{\! d}$ contain electroweak symmetry
breaking doublets. For instance, we have the first order expansions
\beqa
 \Phi_1^\dag &=& \left( \begin{array}{cccc}
0 & 0 & f_{\!\scriptscriptstyle 1} & 0 \end{array} \right) 
+  \frac{i}{\sqrt{2}} \, \frac{ f_{\!\scriptscriptstyle 2}}{ f_{\!\scriptscriptstyle 12}}
 \left( \begin{array}{cccc}
h_{\scriptscriptstyle u}  & 0 & 0 \end{array} \right)
\nonumber \\
 \Psi_2^\dag &=& \left( \begin{array}{cccc}
0 & 0 &  0 & f_{\!\scriptscriptstyle 4} \end{array} \right) 
-  \frac{i}{\sqrt{2}} \, \frac{ f_{\!\scriptscriptstyle 3}}{ f_{\!\scriptscriptstyle 34}}
 \left( \begin{array}{cccc}
h_{\scriptscriptstyle d}  & 0 & 0 \end{array} \right) \;,
\eeqa
with similar expressions for  $\Phi_2^\dag$ and  $\Psi_1^\dag$; where
$f_{\!\scriptscriptstyle 12}^2=f_{\!\scriptscriptstyle 1}^2+f_{\!\scriptscriptstyle 2}^2$
and 
$f_{\!\scriptscriptstyle 34}^2=f_{\!\scriptscriptstyle 3}^2+f_{\!\scriptscriptstyle 4}^2$,
and $h_{\scriptscriptstyle u}$ and $h_{\scriptscriptstyle d}$ are two SM Higgs
doublets. The essential feature here is the vacuum misalignment between the $\Phi_i$
pairs and the $\Psi_i$ pairs. To keep this feature, the third and fourth states of the four
scalar quadruplets must have the same vanishing hypercharge. Enforcing that is
equivalent to enforcing the corresponding choice of the fermionic spectrum, {\it i.e.}
with the fourth state in $Q^a$ being the also top-like (the $T'$). The little Higgs picture
needs a collective symmetry breaking, with aligned  VEVs for the $\Phi_i$, as well
as $\Psi_i$, pair of Higgs multiplets. Beyond that, the existence of both aligned pair
with generic, hence misaligned VEVs actually yields naturally the preserved $U(1)$
symmetry, to be identified as the hypercharge, under which the two states in the
directions of the broken symmetries have vanishing charges.  The choice of fermionic
spectrum with three among the four states of a quadruplet having the same electric
charge is hence more natural than it may look naively. So, the question is whether this
is the only option for little Higgs idea to work.

From the above discussion, we can see that to take a different choice of the fermionic
spectrum, we will have to change the content of the set of the symmetry breaking Higgs
quadruplets as given by Eq.(\ref{phi}). To preserve the VEV misalignment Ref.\cite{KS} 
rely on to fix the (SM)Higgs quartic coupling, the fourth states of the $\Psi_i$ pair 
have to be ones with zero hypercharge. This can be fixed by a suitable choice of their 
$X$-charges. The latter now has to be different from that of the $\Phi_i$ pair.  
Let us explore the possibility with the fermion spectrum of Table III. We still take 
the $\Phi_i$ pairs with $X$-charge
given by $\frac{-1}{2}$ to take care of the top-sector Yukawa couplings.  Added to that, 
we can take a $\Psi_i$ pair with $X$-charge given by $\frac{1}{2}$. The little Higgs
mechanism together with the quartic coupling part as discussed in Ref.\cite{KS}
should still work. Just like the original scenario, each of the four scalar 
quadruplets is related to a global $SU(4)$ symmetry that is broken to $SU(3)$ by
its own VEV. Each pair gives rise to a SM Higgs doublet as pseudo-Nambu-Goldstone
states of a collective breaking of the corresponding pair of $SU(4)$'s.
The spectrum choice, however, dictates a different embedding of 
the SM gauge group and hence gives the $h_{\scriptscriptstyle d}$ doublet a 
different hypercharge. In the case discussed above, and in 
Ref.\cite{KS}, $h_{\scriptscriptstyle d}$ actually has the same 
hypercharge as $h_{\scriptscriptstyle u}$ and does not couple
to the down-sector quarks the way the latter couples to the up-sector, hence a somewhat 
abuse of notation. For the alternative case at hand, however, $h_{\scriptscriptstyle d}$ 
will be an $h_{\scriptscriptstyle d}$ literally. In fact, we have
\beqa 
{\mathcal L}_{bottom} &=& y_{\!\scriptscriptstyle 1}\,\bar{b}_a\,
\Psi_{\!\scriptscriptstyle 1} \, Q^a   +    y_{\!\scriptscriptstyle 2}\,\bar{B}_a\,
\Psi_{\!\scriptscriptstyle 2} \, Q^a
\nonumber \\  &=&
(f_{\!\scriptscriptstyle 3}\,  y_{\!\scriptscriptstyle 1}\,\bar{b}_a + 
f_{\!\scriptscriptstyle 4}\, y_{\!\scriptscriptstyle 2}\,\bar{B}_a)\, B^a
+ \frac{i}{\sqrt{2}} \, \left[ \frac{ f_{\!\scriptscriptstyle 4}}{ f_{\!\scriptscriptstyle 34}}
\, y_{\!\scriptscriptstyle 1}\,\bar{b}_a
-   \frac{ f_{\!\scriptscriptstyle 3}}{ f_{\!\scriptscriptstyle 34}}
\, y_{\!\scriptscriptstyle 2}\,\bar{B}_a \right]\,h_{\scriptscriptstyle d}
\left( \begin{array}{c} t^a \\ b^a \end{array} \right) + \cdots
\label{tykw}
\eeqa
in exact analog to the top sector Yukawa couplings (note: $\bar{b}$ and $\bar{B}$
do not denote exact Dirac partners of $b$ and $B$ here).

At the electroweak scale, the models discussed have two Higgs doublets (plus some
singlets). From the phenomenological point of view, two Higgs doublet models are
prefered to have natural flavor conservation. There is also some indications that
a large $\tan\!\beta$ value is required\cite{2hdm}. The latter implies large bottom
Yukawa coupling. In that case, one does have to worry about the bottom loop 
contribution to the Higgs mass quadratic divergence. The Higgs structure of the
model spectrum given by Table III discussed above could be a little Higgs model
satisfying such requirements. The $b$-$B$ quark pair has canceled quadratic divergent
contributions in exactly the same way as that of the $t$-$T$ quark pair. Hence,
we consider the model to be an interesting alternative little Higgs model. A possible
disadvantage of the model compared to the one above is that, unlike the latter (see 
Ref.\cite{009} for discussions of an analog case) the gauge quantum number
assignments do not rule out renormalizable tree-level Yukawa couplings to SM 
quarks of the first two families. We hope to go into more detailed studies of the
models discussed here in future publications.

To summarize, we have presented here how to construct SM extensions with viable 
fermionic spectra under an $SU(3)_C\times SU(N)_L\times U(1)_X$ gauge symmetry.
The construction methodology is based on a family non-universal treatment, under
which the gauge anomalies cancel among the three SM families in a nontrivial
fashion. While such models could be of interested in their own right, our study is
motivated by solving the hierarchy problem through the little Higgs mechanism.
We also discussed the relevancy and basic features of some of such models as 
little Higgs models, with particular focus on two specific $SU(4)_L$ models.  We
believe that the phenomenology of the models  should be studied in more details.

Our work is partially supported by the National Science Council of Taiwan, under
grant numbers NSC 91-2112-M-008-042 and NSC 92-2112-M-044.


\clearpage
\noindent
Table I : Fermion spectrum for the $SU(3)_C\times SU(3)_L\times U(1)_X$ model with 
little Higgs. Here, we give the hypercharges of the electroweak states, with SM doublets 
put in [.]'s. The other states are singlets. The normalization convention is 
${\mathcal Q}=T_3 +Y$. Hence, singlets have hypercharges identical with electric
charge ${\mathcal Q}$. Other symbols are self-explanatory. Note that despite the 
notation, the fermions are not mass eigenstates. Mass mixings are expected. 
\\
\begin{center}
\begin{tabular}{|c|cc|}
\hline\hline
				& \multicolumn{2}{|c|}{$U(1)_Y$-states}			\\  \hline
${\bf (3_{\scriptscriptstyle C},3_{\scriptscriptstyle L} ,\frac{1}{3})}$                 &  ${\bf \frac{1}{6}}$[$Q$]               & ${\bf \frac{2}{3}}$($T$)       	\\
2\ ${\bf ({3}_{\scriptscriptstyle C},\bar{3}_{\scriptscriptstyle L} ,0)}$     &  2\ ${\bf \frac{1}{6}}$[2\ $Q$]               & 2\ ${\bf \frac{-1}{3}}$($D,S$)     \\
$3\ {\bf (l_{\scriptscriptstyle C} ,3_{\scriptscriptstyle L} ,\frac{-1}{3})}$           & 3\  ${\bf \frac{-1}{2}}$[3\ $L$]          	& 3\ {\bf 0}(3\ $N$)   \\
$4\ {\bf (\bar{3}_{\scriptscriptstyle C},1_{\scriptscriptstyle L} ,\frac{-2}{3})}$        & \multicolumn{2}{|c|}{4\ ${\bf \frac{-2}{3}}$ ($\bar{u}, \bar{c}, \bar{t}, \bar{T}$)} 	     	\\
$5\ {\bf (\bar{3}_{\scriptscriptstyle C},1_{\scriptscriptstyle L} ,\frac{1}{3})}$        & \multicolumn{2}{|c|}{5\ ${\bf \frac{1}{3}}$ ($\bar{d}, \bar{s}, \bar{b}, \bar{D}, \bar{S}$)} 	                 	\\
$3\ {\bf (1_{\scriptscriptstyle C},1_{\scriptscriptstyle L} ,1)}$              &    \multicolumn{2}{|c|}{3\ ${\bf 1}$  ($e^+, \mu^+, \tau^+$)   }      	\\
\hline\hline
\end{tabular}
\end{center}

\bigskip
\bigskip
\bigskip
\bigskip

\noindent
Table II : Fermion spectrum for a  $SU(3)_C\times SU(4)_L\times U(1)_X$ model 
with little Higgs. Again, we give the hypercharges of the electroweak states, with 
SM doublets put in [.]'s. Basic notation is the same as that of Table I. 
Note that we separate in the last column a set of singlet quarks and leptons to
which alternative choices may be a feasibility. In that case, one has to made
adjustments to the some of the $U(1)_X$-charges, as discussed in the text
(see also Table III).
\\
\begin{center}
\begin{tabular}{|c|ccc|}
\hline\hline
				& \multicolumn{3}{|c|}{$U(1)_Y$-states}			\\  \hline
${\bf (3_{\scriptscriptstyle C},4_{\scriptscriptstyle L} ,\frac{5}{12})}$                 &  ${\bf \frac{1}{6}}$[$Q$]               & ${\bf \frac{2}{3}}$($T$)  &      ${\bf \frac{2}{3}}$($T'$) 	\\
2\ ${\bf ({3}_{\scriptscriptstyle C},\bar{4}_{\scriptscriptstyle L} ,\frac{-1}{12})}$     &  2\ ${\bf \frac{1}{6}}$[2\ $Q$]               & 2\ ${\bf \frac{-1}{3}}$($D,S$)     & 2\ ${\bf \frac{-1}{3}}$($D',S'$) \\
$3\ {\bf (l_{\scriptscriptstyle C} ,4_{\scriptscriptstyle L} ,\frac{-1}{4})}$           & 3\  ${\bf \frac{-1}{2}}$[3\ $L$]          	& 3\ {\bf 0}(3\ $N$) 		 & 3\ {\bf 0}(3\ $N'$)   \\
$5\ {\bf (\bar{3}_{\scriptscriptstyle C},1_{\scriptscriptstyle L} ,\frac{-2}{3})}$        & \multicolumn{2}{|c}{4\ ${\bf \frac{-2}{3}}$ ($\bar{u}, \bar{c}, \bar{t}, \bar{T}$)} 	
& { ${\bf \frac{-2}{3}}$ ($\bar{T'}$)} 	          	\\
$7\ {\bf (\bar{3}_{\scriptscriptstyle C},1_{\scriptscriptstyle L} ,\frac{1}{3})}$        & \multicolumn{2}{|c}{5\ ${\bf \frac{1}{3}}$ ($\bar{d}, \bar{s}, \bar{b}, \bar{D}, \bar{S}$)} 	
& 2\ ${\bf \frac{1}{3}}$ ($\bar{D'}, \bar{S'}$)                  	\\
$3\ {\bf (1_{\scriptscriptstyle C},1_{\scriptscriptstyle L} ,1)}$              &    \multicolumn{2}{|c}{3\ ${\bf 1}$  ($e^+, \mu^+, \tau^+$)   }      	 &   	\\
\hline\hline
\end{tabular}
\end{center}

\clearpage

\bigskip\bigskip

\noindent
Table III : Fermion spectrum for another  $SU(3)_C\times SU(4)_L\times U(1)_X$ model 
with little Higgs. Again, we give the hypercharges of the electroweak states, with 
SM doublets put in [.]'s. Basic notation is the same as that of Table II. 
\\
\begin{center}
\begin{tabular}{|c|ccc|}
\hline\hline
				& \multicolumn{3}{|c|}{$U(1)_Y$-states}			\\  \hline
${\bf (3_{\scriptscriptstyle C},4_{\scriptscriptstyle L} ,\frac{1}{6})}$                 &  ${\bf \frac{1}{6}}$[$Q$]               & ${\bf \frac{2}{3}}$($T$)  &      ${\bf \frac{-1}{3}}$($B$) 	\\
2\ ${\bf ({3}_{\scriptscriptstyle C},\bar{4}_{\scriptscriptstyle L} ,\frac{1}{6})}$     &  2\ ${\bf \frac{1}{6}}$[2\ $Q$]               & 2\ ${\bf \frac{-1}{3}}$($D,S$)     & 2\ ${\bf \frac{2}{3}}$($U,C$) \\
$3\ {\bf (l_{\scriptscriptstyle C} ,4_{\scriptscriptstyle L} ,\frac{-1}{2})}$           & 3\  ${\bf \frac{-1}{2}}$[3\ $L$]          	& 3\ {\bf 0}(3\ $N$) 		 & 3\ {\bf -1}(3\ $E^-$)   \\
$6\ {\bf (\bar{3}_{\scriptscriptstyle C},1_{\scriptscriptstyle L} ,\frac{-2}{3})}$        & \multicolumn{2}{|c}{4\ ${\bf \frac{-2}{3}}$ ($\bar{u}, \bar{c}, \bar{t}, \bar{T}$)} 	
& {2\ ${\bf \frac{-2}{3}}$ ($\bar{U}, \bar{C}$)} 	          	\\
$6\ {\bf (\bar{3}_{\scriptscriptstyle C},1_{\scriptscriptstyle L} ,\frac{1}{3})}$        & \multicolumn{2}{|c}{5\ ${\bf \frac{1}{3}}$ ($\bar{d}, \bar{s}, \bar{b}, \bar{D}, \bar{S}$)} 	
& ${\bf \frac{1}{3}}$ ($\bar{B}$)                 	\\
$6\ {\bf (1_{\scriptscriptstyle C},1_{\scriptscriptstyle L} ,1)}$              &    \multicolumn{2}{|c}{3\ ${\bf 1}$  ($e^+, \mu^+, \tau^+$)   }      	 & 3\ {\bf 1}(3\ $E^+$)   	\\
\hline\hline
\end{tabular}
\end{center}


\begin{thebibliography}{99}
\bibitem{dd}
N. Arkani-Hamed, A.G.~Cohen, and H.~Georgi, 
Phys. Rev. Lett. {\bf 86}, 4757 (2001).
\bibitem{S}
See, for a recent review, M. Schmaltz, hep-ph/0210415,
and references therein.
\bibitem{scp}
H.-Y.~Cheng, Phys. Rep. {\bf 158}, 1 (1988).
See, however, T.~Lee, Int. J. Mod. Phys. {\bf A16}, 4321 (2001).
\bibitem{unc67}
O.C.W. Kong, Mod. Phys. Lett. {\bf A11}, 2547 (1996);
Phys. Rev. {\bf D55}, 383 (1997).
\bibitem{SVS}
M. Singer,   J.W.F. Valle, and J. Schechter,  Phys. Rev. {\bf D22}, 738 (1980).
\bibitem{331}
F. Pisano and V. Pleitez,   Phys. Rev. {\bf D46}, 410 (1992);
P.H. Frampton,  Phys. Rev. Lett. {\bf 69}, 2889 (1992).
\bibitem{FLT}
R.~Foot, H.N.~Long, and T.A.~Tran,  Phys. Rev. {\bf D50}, R34 (1994).
\bibitem{unifi}
For examples of unification models extending on $SU(5)$, see
H. Georgi,  Nucl. Phys. {\bf B156}, 126 (1979);
P.H.~Frampton and S.~Nandi, Phys. Rev. Lett. {\bf 43}, 1460 (1979).
\bibitem{unc}
For examples of horizontal symmetry models with or without (vertical) unification
see P.H.~Frampton and O.C.W.~Kong, Phys. Rev. Lett. {\bf 77}, 1699 (1996),
and P.H.~Frampton and O.C.W.~Kong, Phys. Rev. Lett. {\bf 75}, 781 (1995),
respectively.
\bibitem{009}
O.C.W. Kong, hep-ph/0307250, NCU-HEP-k009
\bibitem{KS}
D.E. Kaplan and M. Schmaltz, hep-ph/0302049.
\bibitem{2hdm}
See for example,
K. Cheung, C.H. Chou, and O.C.W. Kong, Phys. Rev. {\bf D64}, {\it 111301(R)}, (2001) ;
K.~Cheung and O.C.W.~Kong, hep-ph/0302111, Phys. Rev. D (2003) {\it (to be published)}.

\end{thebibliography}
\end{document}